\documentclass[conference]{IEEEtran}
\IEEEoverridecommandlockouts
\usepackage{verbatim}%for the comment
\usepackage[utf8x]{inputenc}
\usepackage{epstopdf}
\usepackage{graphicx}
\usepackage{tabularx}
\usepackage{booktabs}
\usepackage[table]{xcolor}
\usepackage{booktabs}
\usepackage{multirow}
\usepackage{ctable}
\usepackage[cmex10]{amsmath}
\usepackage{algorithm}
\usepackage{algpseudocode}
\usepackage{pifont}
\usepackage{amssymb}
\usepackage{cite}
\usepackage{subcaption}
\usepackage{textcomp}
\usepackage{textpos}
\usepackage{array}
\newcommand*\xbar[1]{%
	\hbox{%
		\vbox{%
			\hrule height 0.5pt % The actual bar
			\kern0.5ex%         % Distance between bar and symbol
			\hbox{%
				\kern-0.1em%      % Shortening on the left side
				\ensuremath{#1}%
				\kern-0.1em%      % Shortening on the right side
			}%
		}%
	}%
}

\newcommand{\F}{\operatorname{FS}}
\newcommand{\U}{\operatorname{UE}}
\begin{document}
\bstctlcite{IEEEexample:BSTcontrol}
%---------------------------------------------------------------------------------
%                                    Title
%---------------------------------------------------------------------------------
\title{Coexistence of 5G mmWave Users with Incumbent Fixed Stations over 70 and 80 GHz} %The title of the paper
\author{\IEEEauthorblockN{Ghaith Hattab, Eugene Visotsky, Mark Cudak, and Amitava Ghosh }
	\IEEEauthorblockA{Nokia Bell Labs, Arlington Heights, IL, USA\\
		\{ghaith.hattab, eugene.visotsky, mark.cudak, amitava.ghosh\}@nokia.com}
}

\maketitle

\begin{textblock*}{180mm}(.01\textwidth,-6.2cm)
G. Hattab, E. Visotsky, M. Cudak, and A. Ghosh, “Coexistence of 5G mmWave Users with Incumbent Fixed Stations over 70 and 80 GHz", IEEE GLOBECOM’17, Dec. 2017
\end{textblock*}
\vspace{-.2in}
%---------------------------------------------------------------------------------
%                                    Abstract
%---------------------------------------------------------------------------------
\begin{abstract}
Millimeter wave spectrum access over the 70GHz and 80GHz is central to unlocking gigabit connectivity and meeting the explosive growth of mobile traffic. A pressing question, however, is whether fifth-generation (5G) systems can harmoniously coexist with the incumbents of these bands, which are primarily point-to-point fixed stations (FSs). To this end, we thoroughly analyze the impact of 5G coexistence on FSs. Specifically, we first analyze the geometry of existing FSs' deployment using actual databases of these stations. Then, we present a case study on the interference generated from users towards FSs in two populated areas in Chicago, where we use actual building databases to accurately compute the aggregate interference. The analysis and simulation results reveal that the deployment strategy of FSs and the high attenuation losses at 70/80GHz significantly limit the 5G interference, with the majority of FSs experiencing interference levels well below the noise floor.
\end{abstract}

%---------------------------------------------------------------------------------
%                                    Index Words
%---------------------------------------------------------------------------------

%---------------------------------------------------------------------------------
%                             I. Introduction
%---------------------------------------------------------------------------------
\section{Introduction}
Millimeter wave (mmWave) spectrum access has become a defining feature for fifth-generation (5G) cellular networks \cite{Xiao2017,ITU2015}. While access beyond the sub-6 GHz is not new, it has been only recently considered as a key disruptive feature for cellular networks. Indeed, 5G promises to meet the explosive growth of mobile traffic and to provide unparalleled network capacity, with peak data rates reaching tens of Gbps \cite{AndrewsZhang2014a}. Hence, it is no longer sufficient to enhance spectral efficiency in the sub-6 GHz, and the need for more spectrum has become crucial to scale with the ever-increasing data demands.

The recent interest on mmWave access has led the Federal Communications Commission (FCC) to open up 3.85GHz of licensed spectrum for cellular services, and specifically at 28GHz
(27.5-28.35GHz) and 39GHz (37-40GHz) \cite{FCC2016b}, with major mobile operators, e.g., AT\&T and Verizon, already acquiring licenses in these bands. Nevertheless, there is still an additional 10GHz of licensed spectrum at 70GHz (71-76GHz) and 80GHz (81-86GHz) that are left for future consideration as candidate bands for mmWave mobile networks \cite{FCC2016b}. 

The advantages of using 70 and 80GHz bands, also known as the \emph{e-band}, are twofold. First, each band can easily provide a contiguous high bandwidth, e.g., 2GHz, in contrast to 28GHz and 39GHz, where each provides a maximum of 850MHz and 1.6GHz, respectively. Second, the e-band is available worldwide, enabling economies of scale through universal adoption of common mmWave devices. Equally important, operating at the higher end of the mmWave spectrum is not significantly different from operating at 28GHz as the channel models are the same, and the increase in path loss can be compensated by using an array with a larger number of antenna elements. While Nokia has already demonstrated the feasibility of mmWave systems at 70GHz \cite{Cudak2014,Inoue2017}, a key pressing issue is the coexistence of 5G systems with the incumbents, which are primarily fixed stations (FSs) that provide point-to-point services, e.g.,  wireless backhaul. A preliminary study on such coexistence is presented in \cite{Kim2017}. Nevertheless, it considers a single FS and assumes a predetermined portion of interfering links to be non-line-of-sight (NLOS). In this paper, however, we aim to consider a realistic deployment of FSs to accurately analyze the impact of the aggregate interference generated from 5G uplink (UL) transmissions of user equipment terminals (UEs).

The contributions of this paper are twofold. First, we thoroughly analyze the geometry of incumbents' deployment by parsing actual databases of FSs over four major metropolitan areas in the United States. Such analysis provides key insights on the spatial distribution of these stations, their orientation and pointing directions, and the likelihood of being victims of strong interference from UEs operating in the UL. Second, we present a case study on the interference generated from randomly dropped UEs outdoors in two densely populated areas: Lincoln Park and Chicago Loop. Unlike the work in \cite{Kim2017}, which considers a statistical blockage model, we use an actual building database to determine a blockage event. Our analysis and results reveal that the geometry of existing FSs' deployment significantly limits the opportunity for 5G UE interference. Specifically, it is shown that the aggregate interference is well below the noise floor, with the majority of FSs experiencing an interference-to-noise ratio (INR) below $-6$dB. This emphasizes that a harmonious coexistence of 5G systems with incumbent FSs over 70GHz and 80GHz is realizable thanks to the different deployment strategies of these networks and the high attenuation losses at such frequencies.

The rest of the paper is organized as follows. The study of FSs' deployment and the analysis of 5G UE interference are presented in Section \ref{sec:FSdatabase} and Section \ref{sec:interferenceAnalysis}, respectively. Simulation results are presented in  Section \ref{sec:simulations}, and the conclusions are drawn in Section \ref{sec:conclusion}.

%---------------------------------------------------------------------------------
%                         II. System Model and Problem Formulation
%---------------------------------------------------------------------------------
\section{Analysis of FSs Deployment}\label{sec:FSdatabase}
We consider FSs that are registered and deployed in four major metropolitan areas: Chicago, New York, Los Angeles, and San Fransisco, where the database for each one covers an area of radius 300km. Table \ref{tab:FSdatabase} shows the actual number of registered links in these areas as well as the total number of pairs. A link is defined as a two-way communication channel between two FSs, whereas a pair is defined as a link with unique longitude and latitude coordinates of the FSs. Thus, the same pair could have multiple links, each over a different channel in 70GHz and/or 80GHz. 

\begin{table}[!t]
	\caption{Current number of links and pairs in each database}
	\label{tab:FSdatabase}
	\centering
	\begin{tabular}{|l|c|c|}
		\hline
		Database   		&  No. of links & No. of pairs \\\hline
		Chicago			&  1743			& 512\\
		New York		&  5303			& 1685\\
		Los Angeles		&  1013			& 911\\
		San Francisco	&  1892			& 1801\\	
		\hline
	\end{tabular}
\end{table}

\begin{figure}[t!]
	\centering
	\begin{subfigure}[t]{0.5\textwidth}
		\centering
		\includegraphics[width=2.25in]{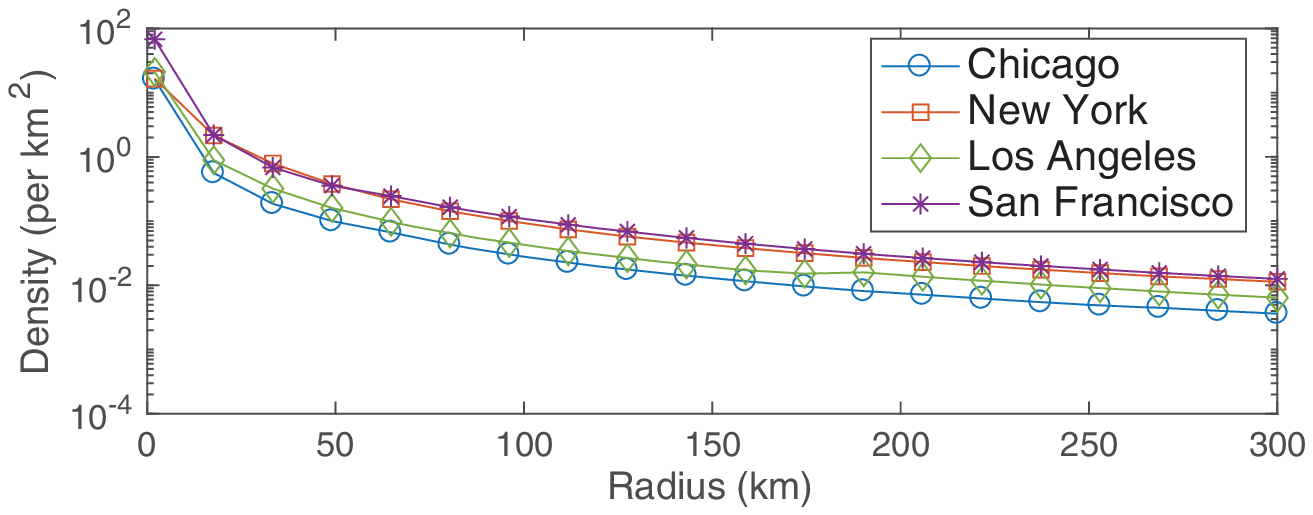}
		\caption{}
		\label{fig:FSdensity}
	\end{subfigure}
	\\
	\begin{subfigure}[t]{.5\textwidth}
		\centering
		\includegraphics[width=2.25in]{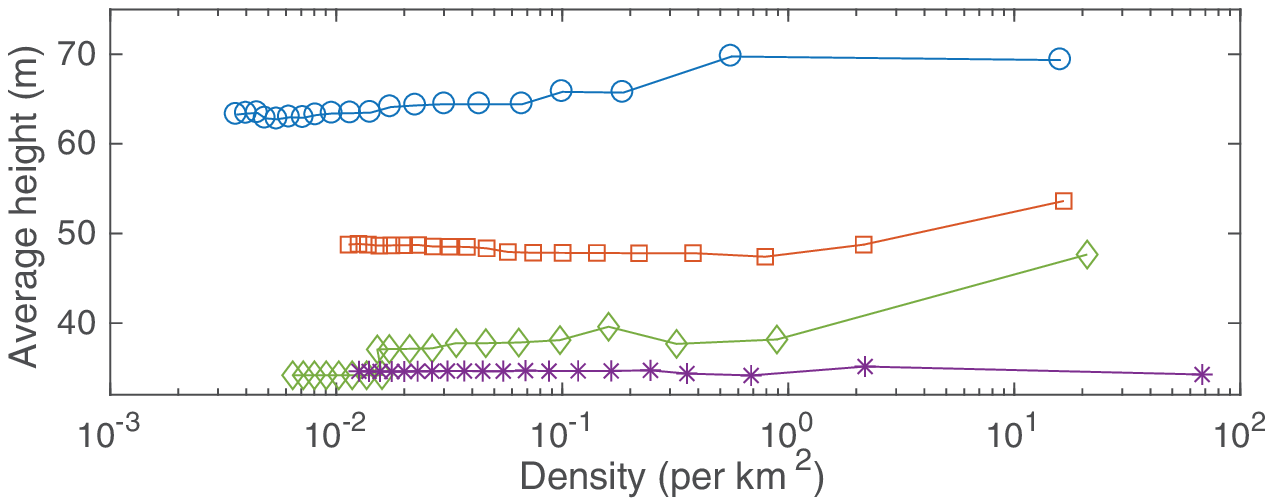}
		\caption{}
		\label{fig:FSheight}
	\end{subfigure}
	\caption{FSs' spatial deployment: (a) Density with variations of region's radius; (b) Average height for a given density.}
	\label{fig:FSdeployment1}
\end{figure}

We first analyze the spatial distribution of these FSs. Fig. \ref{fig:FSdensity} shows their density with variations of the region's radius, where the center of the region is a city center (e.g., Willis Tower for Chicago, the Empire State Building for New York, and the financial districts of Los Angeles and San Fransisco). It is evident that FSs are non-uniformly distributed over space, and specifically they tend to have higher density near city centers while they become very sparsely deployed in suburban areas. Overall, FSs have low density relative to existing cellular networks. Fig. \ref{fig:FSheight} shows the average height of FS deployment for a given density. It is shown that, except for San Francisco, the average height generally increases in denser areas compared to lightly dense areas, showing that the deployment height appears to be correlated with the average building heights in these areas. From the 5G coexistence perspective, this implies that the density of FSs in urban areas should not be worrisome as these stations tend to be deployed at altitudes that are above  5G cell sites. Similarly, in suburban areas, it is more likely to have some FSs at relatively low heights, yet they tend to be sparsely deployed.

\begin{figure}[t!]
	\center
	\includegraphics[width=2.25in]{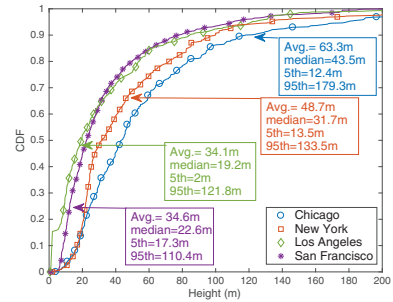} 
	\caption{Height distribution of FSs.}
	\label{fig:heightCDF}
\end{figure}

Fig. \ref{fig:heightCDF} shows the cumulative density function (CDF) of the FSs' deployment height. The average and median heights are at least 34m and 19m, respectively. More importantly, 95\% of FSs are deployed above 12m for most metropolitan areas. Note that for LA, the fifth percentile is 2m, but this is relative to ground, i.e., many of FSs in LA are actually deployed on hills. Since 5G sites are expected to be deployed at heights of four to six meters, 5G base stations (gNBs) will be below the majority of FSs, limiting the 5G interference on FSs and vice versa. 

\begin{figure*}[t!]
	\centering
	\begin{subfigure}[t]{.3\textwidth}
		\centering
		\includegraphics[width=2.4in]{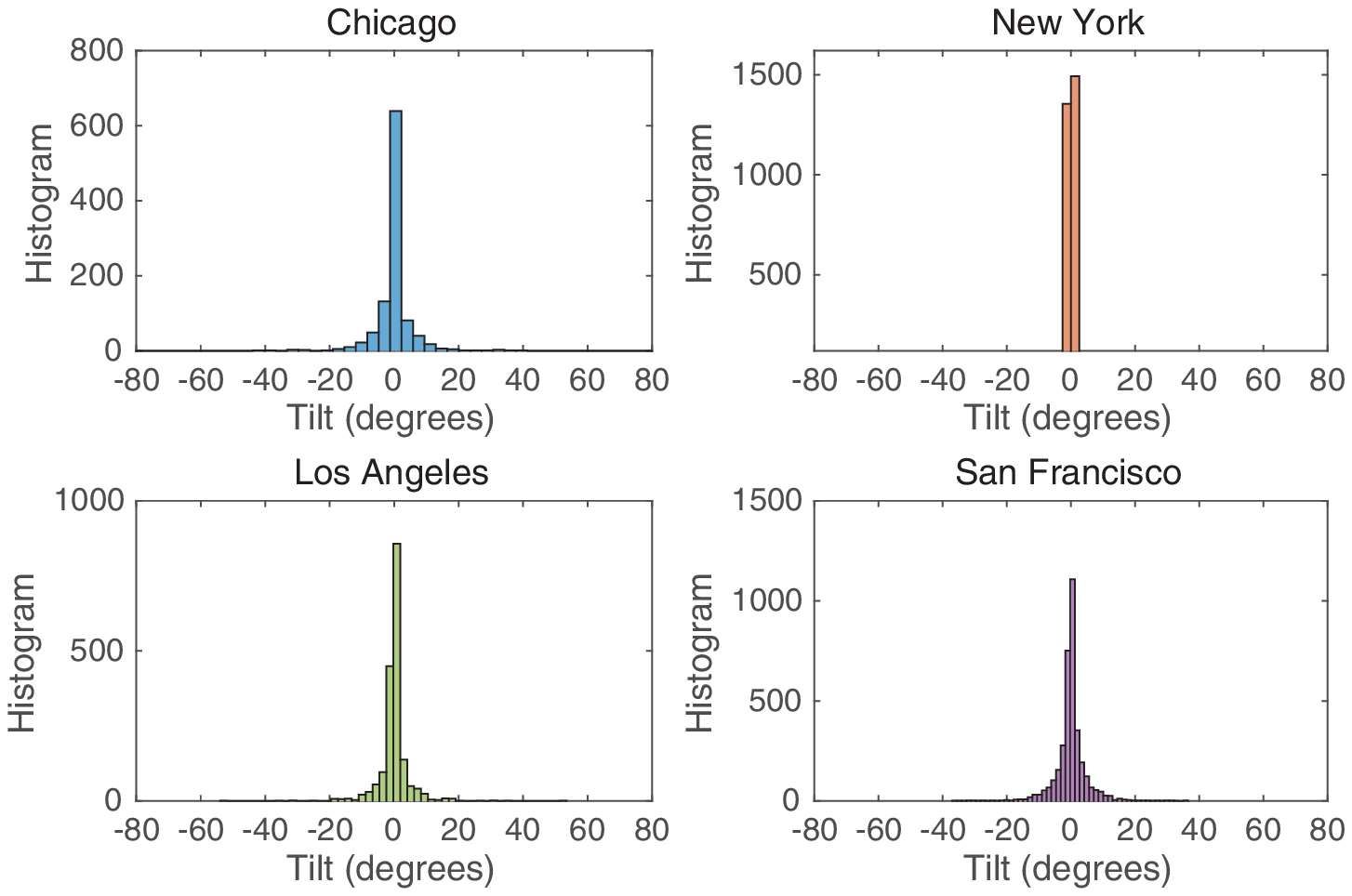}
		\caption{}
		\label{fig:tiltHist}
	\end{subfigure}
	~
	\begin{subfigure}[t]{.3\textwidth}
		\centering
		\includegraphics[width=2.4in]{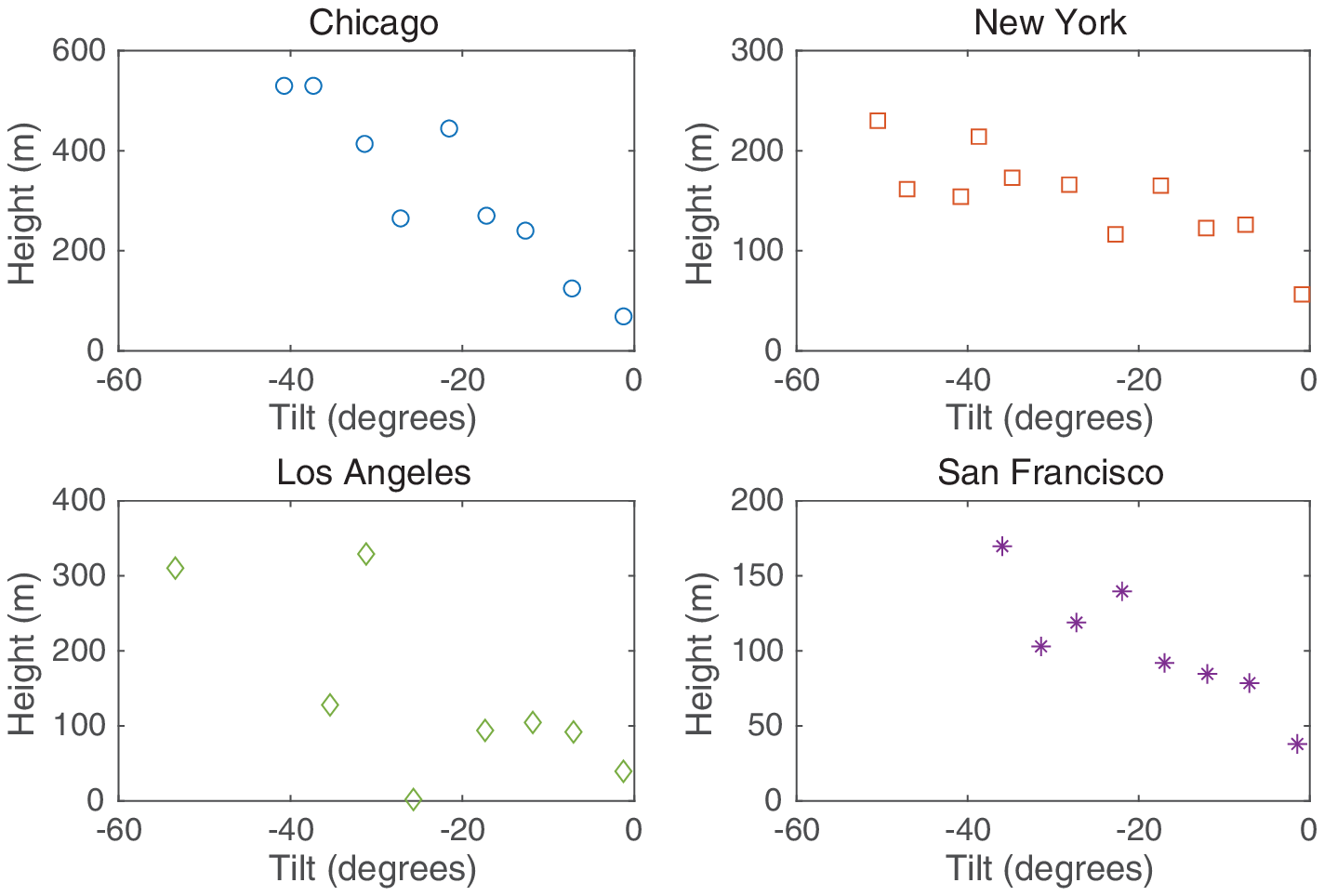}
		\caption{}
		\label{fig:tiltVsHeight}
	\end{subfigure}
	~
	\begin{subfigure}[t]{.3\textwidth}
	\centering
	\includegraphics[width=2.4in]{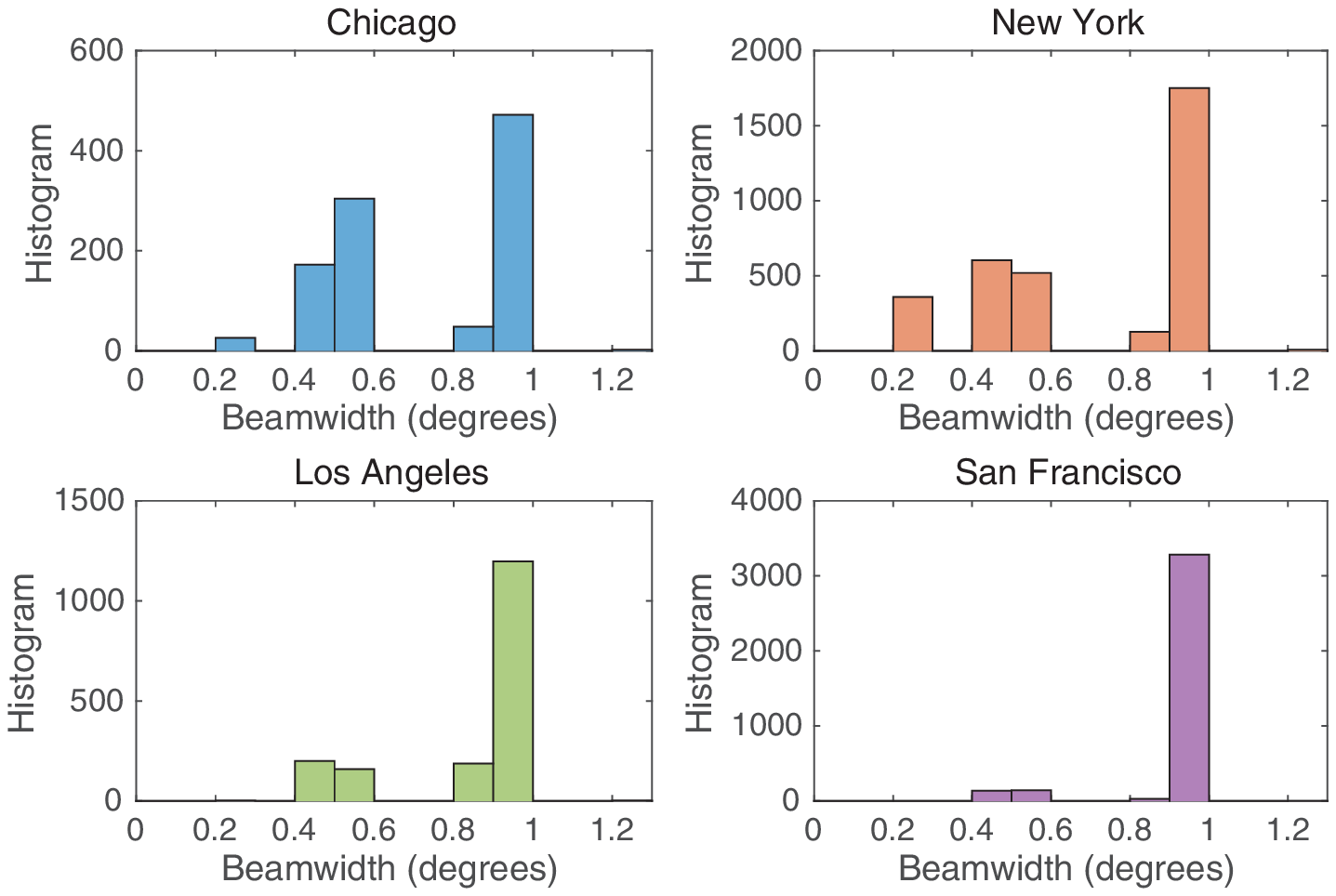}
	\caption{}
	\label{fig:beamwidthHist}
	\end{subfigure}
	\caption{FSs' antenna information: (a) Tilt histograms; (b) average height for a given tilt; (c) beamwidth histograms.}
	\label{fig:FSdeployment2}
\end{figure*}

Another critical aspect of FSs' deployment is their physical antenna orientation. Fig. \ref{fig:tiltHist} shows the histogram of the antenna's tilt, verifying that the vast majority of FSs have their tilt angles pointing horizontally. For instance, more than 93\% of FSs have their tilt angles within $[-10,10]$ degrees. Although there are very few FSs with high negative tilts, i.e., they point to the street level, those FSs are typically deployed at very high altitudes as shown in Fig. \ref{fig:tiltVsHeight}. In other words, there is a correlation between the deployment height and the negative tilt. Thus, although these FSs will have a higher chance to experience UE interference, as they point to the ground, 5G signals will typically see a larger path loss given the height of these FSs. 

Another key feature of FSs is their ultra-narrow beamwidths. Specifically, as per the FCC regulations \cite{FCC2017}, the maximum 3dB beamwidth should be less than or equal to $1.2^{\circ}$. This is verified in Fig. \ref{fig:beamwidthHist}, where the vast majority of FSs have beamwidths at $1^{\circ}$. From a 5G coexistence perspective, the UE must be tightly aligned with the FS for it to cause tangible interference. Otherwise, most 5G signals will be highly attenuated, falling outside the FS receiver's beam. 

In summary, the deployment strategy of FSs is favorable for future 5G deployment over 70GHz and 80GHz for the following reasons.

\begin{itemize}
	\item FSs are generally deployed above 12m, whereas 5G cell sites will be only at 4 to 6 meters above the ground for street level deployment, and hence they will be well below FSs.
	\item The vast majority of FSs are oriented horizontally, i.e., they are directed above 5G deployments. For the few FSs that point to the street level, these are typically at high altitudes, increasing the path loss between the UE and the FS. 
	\item The ultra-narrow beamwidths of FSs can help significantly attenuate UE interfering signals when they fall outside the main lobe. 
\end{itemize}

%---------------------------------------------------------------------------------
%                         III. Analysis of UE Interference on FSs
%---------------------------------------------------------------------------------
\section{Analysis of UE Interference on FSs}\label{sec:interferenceAnalysis}
We focus on 5G UE interference on FSs for the following reasons. First, UEs typically have positive tilt angles compared to 5G gNBs, and thus the former are more likely to interfere with FSs. Second, the mobility of UEs makes their location to appear random, while gNBs' deployment can be optimized to ensure minimal interference on FSs. In addition, we only consider outdoor deployment of UEs, as FSs are outdoors and the attenuation due to penetration losses for indoor UEs is very high at 70GHz and 80GHz. 

The interference seen at a FS is an aggregation of all UEs in vicinity transmitting in the UL to their respective gNBs. Such aggregated interference depends mainly on three components: (i) The path loss between the UE and the FS, (ii) the attenuation due to the FS's antenna pattern, and (iii) the attenuation due to the UE's antenna pattern. We describe each one in details next.

\subsection{Path Loss between a User and a Fixed Station}
We use the 3GPP NR-UMi path loss model \cite{3GPP2017}. Specifically, the path loss between a UE-FS link, in dB, is expressed as
\begin{equation}
\operatorname{PL} = \mathbf{1}_{(\beta=0)} \operatorname{PL}_{\operatorname{LOS}}+\mathbf{1}_{(\beta=1)} \operatorname{PL}_{\operatorname{NLOS}},
\end{equation}
where $\operatorname{PL}_{\operatorname{LOS}}$ is the line-of-sight (LOS) path loss, $\operatorname{PL}_{\operatorname{NLOS}}$ is the non-LOS (NLOS) path loss, $\beta\in\{0,1\}$ is a binary variable that indicates whether the UE-FS is blocked by a building or not, and $\mathbf{1}(\cdot)$ is the indicator function. We emphasize, here, that unlike the 3GPP model in \cite{3GPP2017}, which uses a probabilistic approach to model blockage, we use an actual database to compute the presence/absence of building blockage. We note that both $\operatorname{PL}_{\operatorname{LOS}}$ and $\operatorname{PL}_{\operatorname{NLOS}}$ are functions of the distance between the UE and the FS, their heights, and the center frequency, as given in \cite{3GPP2017}. In addition, both components include log-normal shadow fading, where the LOS and NLOS standard deviations are, respectively, $\sigma_{\operatorname{LOS}}=4$dB and $\sigma_{\operatorname{NLOS}}=7.82$dB \cite{3GPP2017}. 
 
\begin{figure}[t!]
	\center
	\includegraphics[width=2.5in]{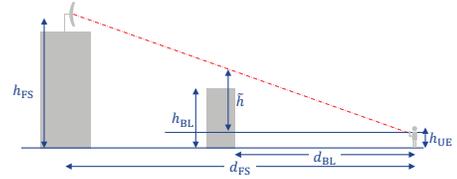} 
	\caption{A blockage event in 3D occurs when $\tilde h+h_{\U}\leq h_{\operatorname{BL}}$.}
	\label{fig:blockageCheck}
\end{figure}

The blockage event is defined as having the UE-FS blocked by a building. This is computed as follows. Assuming the $xy$-plan represents the ground, we first check whether the line that connects between the UE and the FS is blocked by a building in 2D. If the building, defined by a polygon, does intersect with the line, we then check whether it blocks the line in 3D. Specifically, let $d_{\operatorname{BL}}$ be the distance between the UE and the building and $d_{\F}$ be the distance between the UE and the FS. Further, let $h_{\U}$, $h_{\operatorname{BL}}$, and $h_{\F}$, be the heights of the UE, the building, and the FS, respectively. Then, a blockage event occurs if $\tilde h + h_{\U}\leq h_{\operatorname{BL}}$, where
\begin{equation}
\tilde h = d_{\operatorname{BL}} \times \tan \left(\tan^{-1}\left(\frac{h_{\F}-h_{\U}}{d_{\F}}\right)\right).
\end{equation}
This is visualized in Fig. \ref{fig:blockageCheck}.

\subsection{Attenuation due to FS Antenna Pattern}
It is imperative to consider the attenuation resulted from the misalignment between the UE's beam and the FS's beam as the latter is very narrow yet has very high gain. To this end, we define the line connecting the UE to the FS as the \emph{interference axis}. Let the off-axis azimuth angle $\theta_{\operatorname{off}}^{\F}$ be the angle between the FS's beam direction and the interference axis. Thus, if the $(x,y)$ coordinates of the FS transmitter, the FS receiver, and the UE, are $\mathbf{x}_{\operatorname{Tx}}^{\F}$, $\mathbf{x}_{\operatorname{Rx}}^{\F}$, and $\mathbf{x}_{\U}$, respectively, then
\begin{equation}
\theta_{\operatorname{off}}^{\F} = \cos^{-1}\left( \frac{(\mathbf{x}_{\operatorname{Rx}}^{\F}-\mathbf{x}_{\operatorname{Tx}}^{\F})^{\textsf{T}}(\mathbf{x}_{\operatorname{Rx}}^{\F}-\mathbf{x}_{\U})}{\|\mathbf{x}_{\operatorname{Rx}}^{\F}-\mathbf{x}_{\operatorname{Tx}}^{\F}\|\|\mathbf{x}_{\operatorname{Rx}}^{\F}-\mathbf{x}_{\U}\|}\right),
\end{equation}
where $\|\cdot\|$ is the vector's norm. Similarly, let $\phi_{\operatorname{off}}^{\F}$ be the off-axis elevation angle, then it can be shown that
\begin{equation}
\phi_{\operatorname{off}}^{\F} = \phi_{\operatorname{tilt}} + \tan^{-1}\left(\frac{h_{\F}-h_{\U}}{d_{\F}}\right).
\end{equation}
Both off-axis angles are shown in Fig. \ref{fig:FSangles}. We then use the FCC regulations in \cite{FCC2017} for the azimuth and elevation antenna attenuations denoted, respectively, by $A_{az}^{\F}(\theta_{\operatorname{off}}^{\F})$ and $A_{\operatorname{el}}^{\F}(\phi_{\operatorname{off}}^{\F})$. The FCC model is merely a look-up table, where different attenuation is incurred for different angle ranges. Finally, the FS antenna gain is computed as
\begin{equation}
\label{eq:FSpattern}
G_{\F}=
G_{\operatorname{max}}^{\F} - \min\left\{A_{\operatorname{az}}^{\F}(\theta_{\operatorname{off}}^{\F})+A_{\operatorname{el}}^{\F}(\phi_{\operatorname{off}}^{\F}),A_{\operatorname{FTBR}}^{\F}\right\},
\end{equation}
where $G_{\operatorname{max}}^{\F}$ is the maximum antenna gain in dBi and $A_{\operatorname{FTBR}}^{\F}$ is the front-to-back ratio loss (FTBR) in dB.

\subsection{Attenuation due to UE Antenna Pattern}
\begin{figure}[t!]
	\centering
	\begin{subfigure}[t]{.5\textwidth}
		\centering
		\includegraphics[width=2.5in]{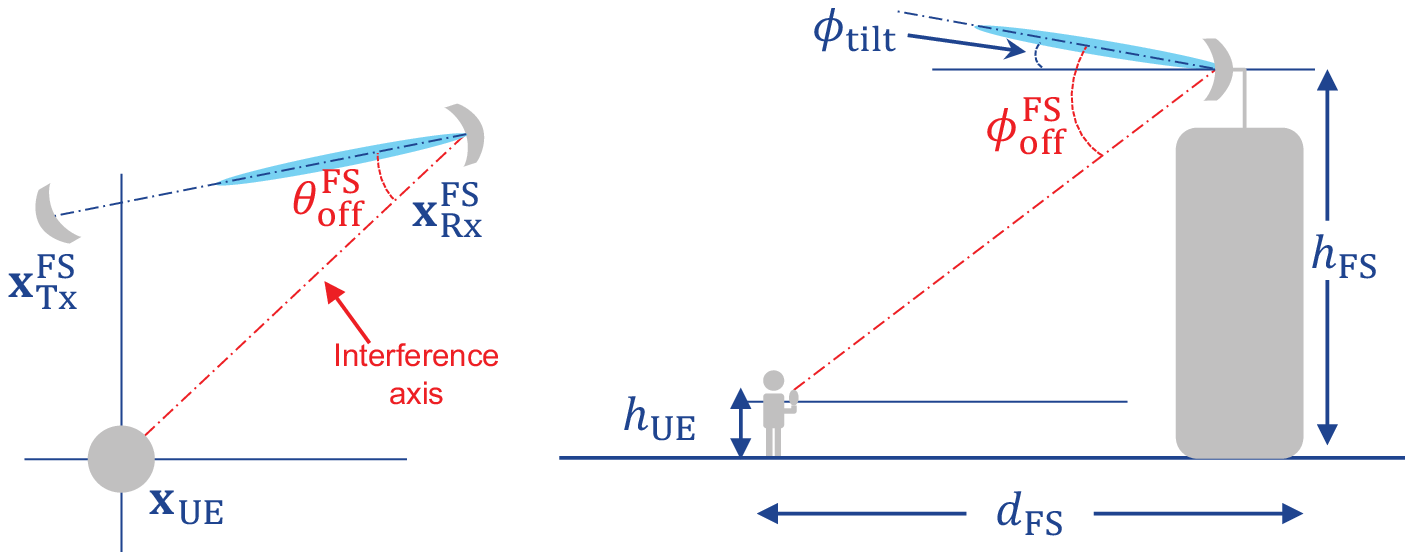}
		\caption{}
		\label{fig:FSangles}
	\end{subfigure}
	\\
	\begin{subfigure}[t]{.5\textwidth}
		\centering
		\includegraphics[width=2.5in]{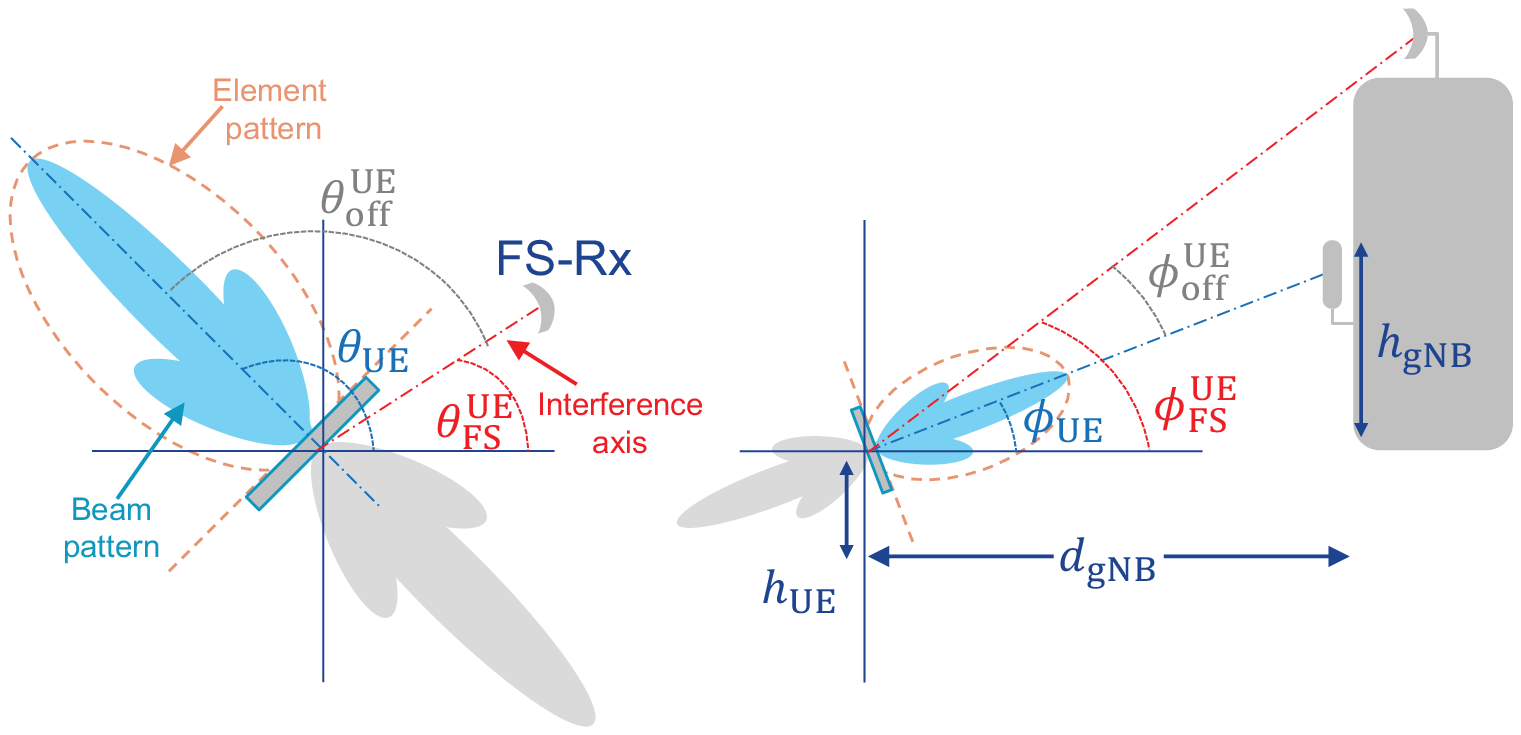}
		\caption{}
		\label{fig:UEangles}
	\end{subfigure}
	\caption{Off-axis azimuth and elevation angles: (a) with respect to the FS; (b) with respect to the UE.}
	\label{fig:AntennaPatterns}
\end{figure}

\begin{figure}[t!]
	\center
	\includegraphics[width=2.5in]{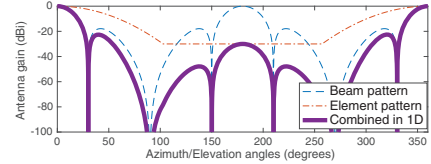} 
	\caption{UE beam and element patterns, assuming $\theta_{\operatorname{UE}}=\phi_{\operatorname{UE}}=0^\circ$.}
	\label{fig:UEpattern}
\end{figure}

We assume that the UE beam direction with respect to the FS is random, where the azimuth direction is uniformly distributed as $\theta_{\U}\sim\mathcal{U}(0,360)$ and the elevation direction is
\begin{equation}
\phi_{\U} = \tan^{-1}\left(\frac{h_{\operatorname{gNB}}-h_{\U}}{d_{\operatorname{gNB}}}\right),
\end{equation} 
where $h_{\operatorname{gNB}}$ is the height of the gNB and $d_{\operatorname{gNB}}\sim\mathcal{U}(10,100)$m is the distance between the UE and the gNB. Thus, the off-axis azimuth and elevation angles are computed as
\begin{equation}
\begin{aligned}
\theta_{\operatorname{off}}^{\U} &= (\theta_{\U}-\theta_{\F}^{\U})\mod 360,\\
\phi_{\operatorname{off}}^{\U}   &= (\phi_{\F}^{\U}-\phi_{\U})\mod 360,
\end{aligned}
\end{equation}
where all angles are illustrated in Fig. \ref{fig:UEangles}. Using these angles, we can compute the beam pattern attenuation in azimuth $A_{\operatorname{BP,az}}^{\U}(\theta_{\operatorname{off}}^{\U})$ and elevation $A_{\operatorname{BP,el}}^{\U}(\phi_{\operatorname{off}}^{\U})$. As for the element pattern, the attenuation in each direction is expressed as
\begin{equation}
A_{\operatorname{EP,az}}^{\U}(\theta_{\operatorname{off}}^{\U})= 12 \left(\frac{\theta_{\operatorname{off}}^{\U}}{\theta_{\operatorname{3dB}}}\right)^2,~~  A_{\operatorname{EP,el}}^{\U}(\phi_{\operatorname{off}}^{\U}) = 12 \left(\frac{\phi_{\operatorname{off}}^{\U}}{\phi_{\operatorname{3dB}}}\right)^2,
\end{equation}
where $\theta_{\operatorname{3dB}}$ and $\phi_{\operatorname{3dB}}$ are the 3dB beamwidth in azimuth and elevation, respectively. Finally, the total signal power radiated from the UE in the direction of the FS is 
\begin{equation}
\begin{aligned}
G_{\U}&= \operatorname{EIRP}_{\operatorname{max}}^{\U} -A_{\operatorname{BP,az}}^{\U}(\theta_{\operatorname{off}}^{\U})-A_{\operatorname{BP,el}}^{\U}(\phi_{\operatorname{off}}^{\U})\\
&- \min\left\{A_{\operatorname{EP,az}}^{\U}(\theta_{\operatorname{off}}^{\U})+A_{\operatorname{EP,el}}^{\F}(\phi_{\operatorname{off}}^{\U}),A_{\operatorname{FTBR}}^{\U}\right\},
\end{aligned}
\end{equation}
where $\operatorname{EIRP}_{\operatorname{max}}^{\U}$ is the UE maximum EIRP and $A_{\operatorname{FTBR}}^{\U}$ is the FTBR loss. Fig. \ref{fig:UEpattern} shows the normalized UE antenna attenuation, where we consider the UE to have 32 antennas using two planar arrays each of size $4\times4$. The 3dB beamwidths of the beam and element patterns are assumed to be $25^{\circ}$ and $65^{\circ}$, respectively, and they are symmetric in azimuth and elevation. 

\subsection{UE Aggregate Interference}
The aggregate 5G UE interference at a given FS is $I_{\operatorname{agg}}=\sum_{i} I_i$, where the $i$-the UE interference is denoted by $I_i$, and it is expressed in dBm as $I_{i,\operatorname{dBm}} = G_{\U}+ G_{\F} - \operatorname{PL}$. In addition, the INR is defined as $\operatorname{INR}_{\operatorname{dB}}= I_{\operatorname{agg,dBm}} - P_{N,\operatorname{dBm}}$, where $P_{N,\operatorname{dBm}}= 10\log_{10}(N_0 B)+F_{\operatorname{dB}}$, where $N_0$ is the noise power spectral density (mW/Hz), $B$ is the bandwidth (Hz), and $F_{\operatorname{dB}}$ is the noise figure of the FS (dB).

%---------------------------------------------------------------------------------
%                         IV. Simulation Results
%---------------------------------------------------------------------------------
\section{Simulation Results}\label{sec:simulations}
\begin{figure}[t!]
	\centering
	\begin{subfigure}[t]{.5\textwidth}
		\centering
		\includegraphics[width=1.85in]{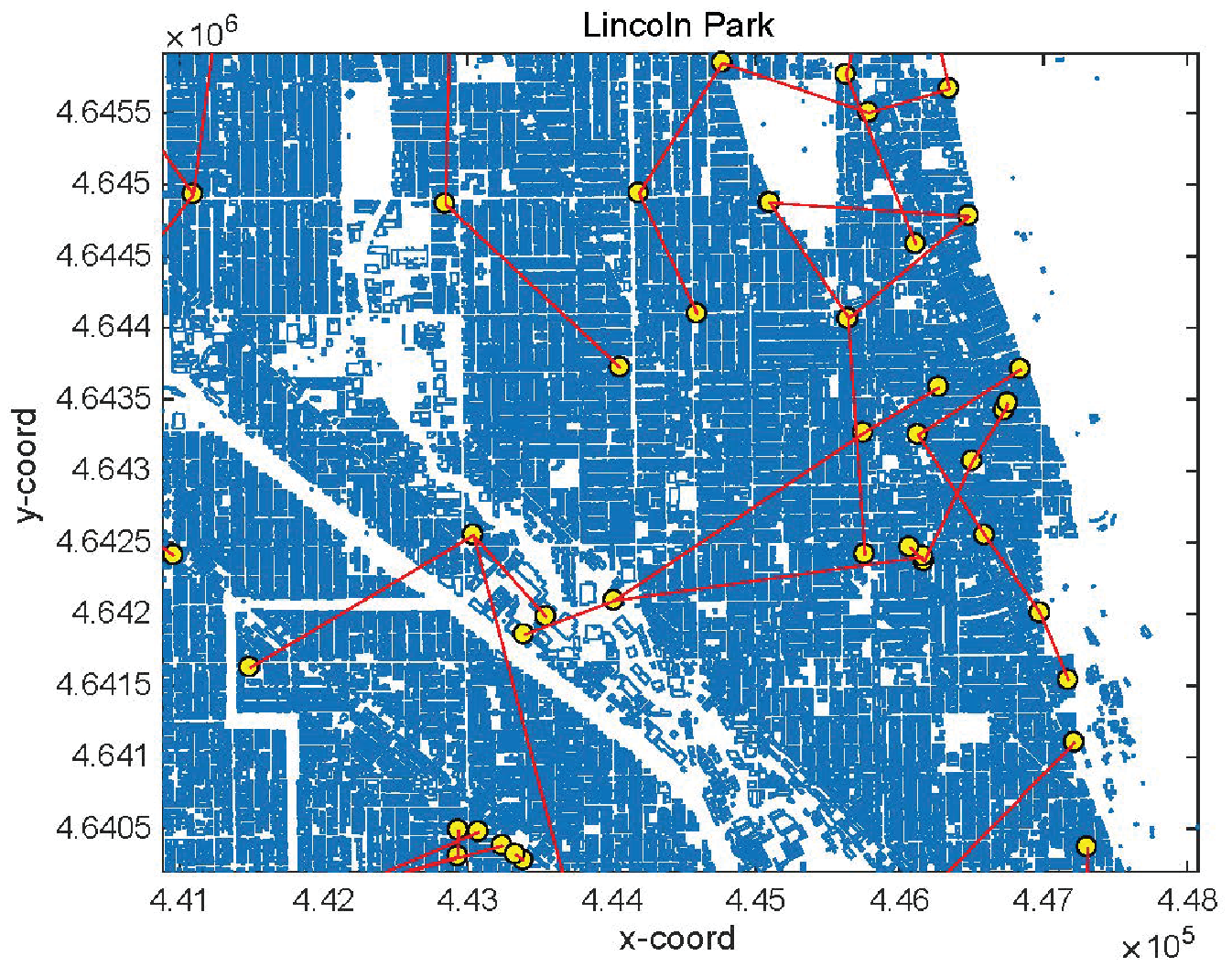}
		\caption{}
	\end{subfigure}
	\\
	\begin{subfigure}[t]{.5\textwidth}
		\centering
		\includegraphics[width=1.85in]{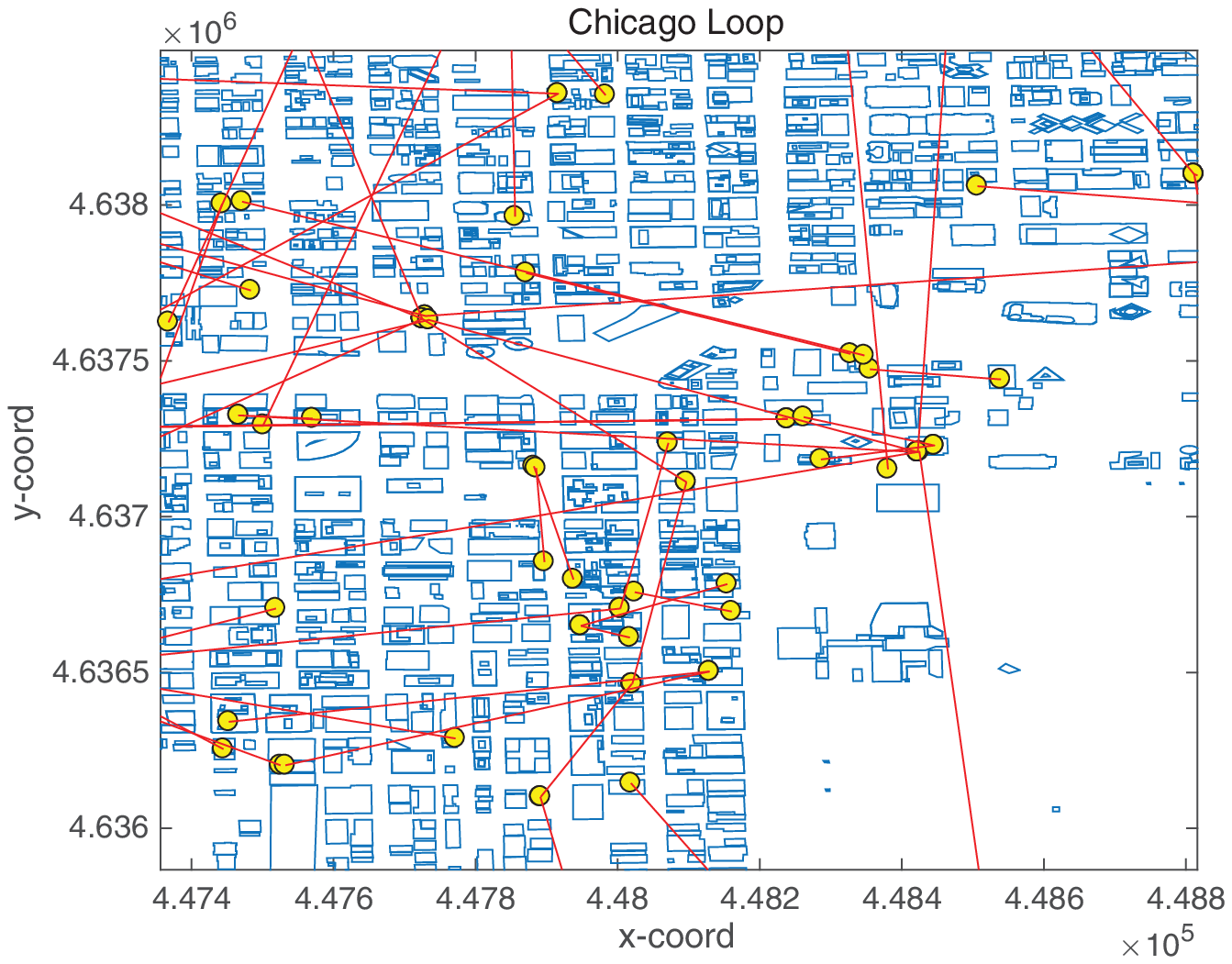}
		\caption{}
	\end{subfigure}
	\caption{(a) Lincoln Park; (b) Chicago Loop.}
	\label{fig:deployment}
\end{figure}

We study the aggregate UE interference on FSs deployed in Lincoln Park and Chicago Loop, which are shown in Fig. \ref{fig:deployment}. It is assumed that the 5G inter-site distance is 200m, and that there are 4 sectors per site. We consider a 25\% instantaneous load in the available UL slots, which translates to 920 and 100 UEs in Lincoln Park and Chicago Loop, respectively. We consider the center frequencies: 73.5GHz and 83.5GHz. We further assume that the UE parameters are $\operatorname{EIRP}_{\operatorname{max}}^{\U}=\{33,43\}$dBm, $A_{\operatorname{FTBR}}^{\U}=30$dB, $h_{\U}=1.5$m, and $h_{\operatorname{gNB}}=6$m. Per FCC regulations, we consider $A_{\operatorname{FTBR}}^{\F}=55$dB \cite{FCC2017}. For noise power, we assume $B=1$GHz and $N_0$ is computed at temperature 290K. Finally, the FS's location, height, maximum antenna gain, antenna tilt, and noise figure, are all extracted from the database. It is assumed that all FSs operate over the same band, as a worst case scenario. The results are averaged out over 100 spatial realizations of UEs.  

\begin{figure}[t!]
	\centering
	\begin{subfigure}[t]{.5\textwidth}
		\centering
		\includegraphics[width=2.65in]{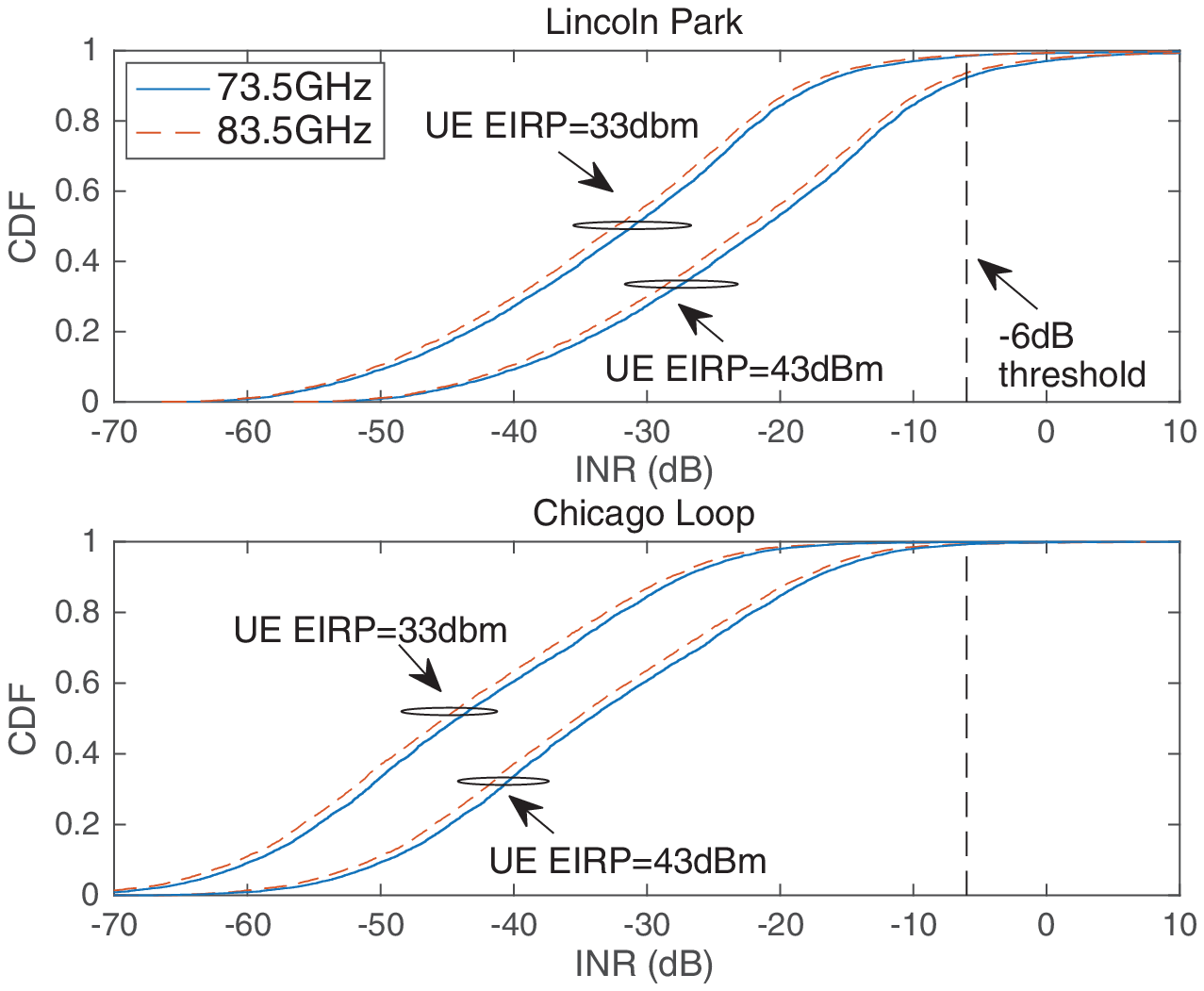}
		\caption{}
		\label{fig:INRcdf}
	\end{subfigure}
	\\
	\begin{subfigure}[t]{.5\textwidth}
		\centering
		\includegraphics[width=2.65in]{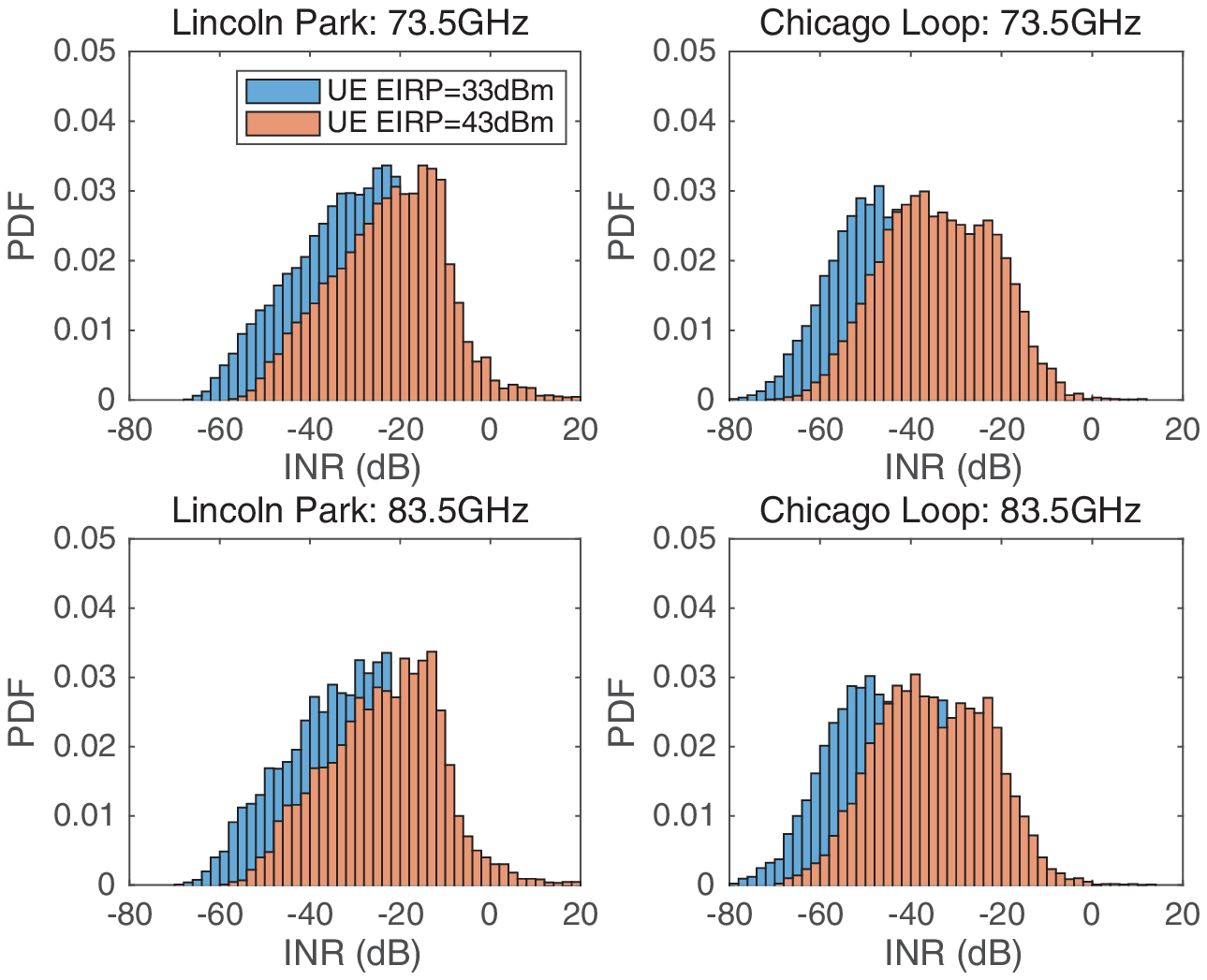}
		\caption{}
		\label{fig:INRpdf}
	\end{subfigure}
	\caption{Distribution of INR: (a) CDF; (b) PDF.}
	\label{fig:INR}
\end{figure}

Fig. \ref{fig:INRcdf} shows the CDF of the INR under different EIRP and operating frequencies. We also show an INR threshold of $-6$dB, which corresponds to a degradation of the signal-to-interference-plus-noise ratio (SINR) of 1dB. Overall, it is evident that the additional UE interference on FSs is well below the noise floor. For instance, the mean INR over 70GHz in Lincoln Park is $-32.0$dB and $-22.1$ for $\operatorname{EIRP}_{\operatorname{max}}^{\U}=33$dBm and $\operatorname{EIRP}_{\operatorname{max}}^{\U}=43$dBm, respectively, while the 95\% percentile is $-13.5$dB and $-3.6$dB. For Chicago Loop, the INR is lower as the number of UEs is less compared to the number of UEs in Lincoln Park, as shown in Table \ref{tab:INRstatistics}. We note that the interference analysis ignores attenuation losses due to vegetation, foliage, cars, and humans \cite{Thomas2014}. Thus, we expect the INR to be even lower due to the presence of these objects. 

Fig. \ref{fig:INRpdf} illustrates the probability density function (PDF) of the INR. It is shown that the majority of FSs are well protected due to the high attenuation at 70GHz and 80GHz as well as the very low likelihood of UEs being aligned within 1$^\circ$ of the FS's beam. Our results have also shown that only a couple of FSs experience INR levels above the $-6$dB interference threshold. These particular FSs are deployed at low heights in areas with wide open space and clear LOS. In this case, simple passive mitigation techniques can be used to limit the interference. For instance, the placement of gNBs can be done such that the 5G beam coverage is perpendicular to the FS, or the 5G gNB can employ exclusion angles to omit beams that correspond to 5G UE beams pointed at the FSs. Such passive mitigation techniques will be the focus of our future work. 

\begin{table}[!t]
	\caption{INR statistics in dB: (70,80) GHz}
	\label{tab:INRstatistics}
	\centering
	\begin{tabular}{|l|>{\centering\arraybackslash}m{0.2in}|>{\centering\arraybackslash}m{0.65in}|>{\centering\arraybackslash}m{0.65in}|>{\centering\arraybackslash}m{0.65in}|}
		\hline
		Statistics   									&EIRP	&  Mean 				&Median 			&95\%				\\\hline
		\multirow{2}{*}{\parbox{1cm}{Lincoln\\Park}}	&33		&  ($-$32.0,$-$33.1)	& ($-$31.1,$-$32.4)	& ($-$13.5,$-$14.5)	\\\cline{2-5}
														&43		&  ($-$22.1,$-$23.2)	& ($-$21.1,$-$22.5) & ($-$3.6,$-$4.6)	\\\cline{1-5}
		\multirow{2}{*}{\parbox{1cm}{Chicago\\Loop}}	&33		&  ($-$43.7,$-$44.8)	& ($-$44.2,$-$45.4)	& ($-$23.9,$-$25.0)	\\\cline{2-5}
														&43		&  ($-$33.7,$-$34.9)	& ($-$34.2,$-$35.5) & ($-$14.0,$-$15.0)	\\\cline{1-5}\hline
	\end{tabular}
\end{table}

%---------------------------------------------------------------------------------
%                         V. Conclusion
%---------------------------------------------------------------------------------
\section{Conclusion}\label{sec:conclusion}
We have analyzed the impact of the aggregate interference generated from 5G UEs in the UL on existing incumbents at 70GHz and 80GHz using actual databases of FSs and buildings. The analysis has shown that the deployment strategy of FSs is favorable for future 5G coexistence as FSs tend to be deployed well above 5G sites, are oriented horizontally, and have narrow beams that are unlikely to be aligned with UEs. In addition, the high propagation losses, particularly due to blockage, at such high frequencies ensure that a typical FS will experience minimal interference from UEs. For the few FSs, that are deployed at low heights in a clear LOS with UEs in vicinity, additional passive mitigation techniques such as the careful placement of gNBs and beam management via exclusion zones, should be sufficient to ensure a harmonious coexistence between 5G systems and incumbents at 70/80GHz.

%---------------------------------------------------------------------------------
%                         References
%---------------------------------------------------------------------------------
\bibliographystyle{IEEEtran}
\bibliography{References}
\end{document}